# Surface Curvature-Induced Directional Movement of Water Droplets

Cunjing Lv[1], Chao Chen[1], Yajun Yin, Quanshui Zheng[2]

*Department of Engineering Mechanics and Center for Nano and Micro Mechanics, Tsinghua University, Beijing 100084, China*

**Abstract**

Here we report a surface curvature-induced directional movement phenomenon, based on molecular dynamics simulations, that a nanoscale water droplet at the outer surface of a graphene cone always spontaneously moves toward the larger end of the cone, and at the inner surface toward the smaller end. The analysis on the van der Waals interaction potential between a single water molecule and a curved graphene surface reveals that the curvature with its gradient does generate the driving force resulting in the above directional motion. Furthermore, we found that the direction of the above movement is independent of the wettability, namely is regardless of either hydrophobic or hydrophilic of the surface. However, the latter surface is in general leading to higher motion speed than the former. The above results provide a basis for a better understanding of many reported observations, and helping design of curved surfaces with desired directional surface water transportation.

---

[1] The two authors Cunjing Lv and Chao Chen have equal contributions to this paper.

[2] To whom correspondence should be addressed. E-mail: zhengqs@tsinghua.edu.cn

**Introduction**

Recently, self-motion phenomena of liquid droplets at solid surfaces have attracted a lot of attentions. [1-5] It was found that silicone oil droplets of volume 3 $\mu$L deposited on the surface of a conical fiber of diameters 0.1 mm spontaneously evolved into annual-shaped, and then moved towards to larger cross-section region, as schematically illustrated in Fig. 1(a) for the droplet outside the conical surface. [6] In contrast, water droplets tapered inside a conical tube were found to move towards the smaller cross-section region, as also illustrated in Fig. 1(a) [7]. Interestingly, glycerol droplets were found to have similar directional motion inside a conical tube [8]. It was further demonstrated that the surface tension can drive annular droplet to move inside a conical tube both theoretically and experimentally. More recently, it was reported [9] that spider silk is capable of collecting water efficiently from the air. The observation showed a similar self-motion mechanism, namely, annular water droplets move at conical surfaces, although the authors noted another mechanism that the wettability may vary with conical axial direction.

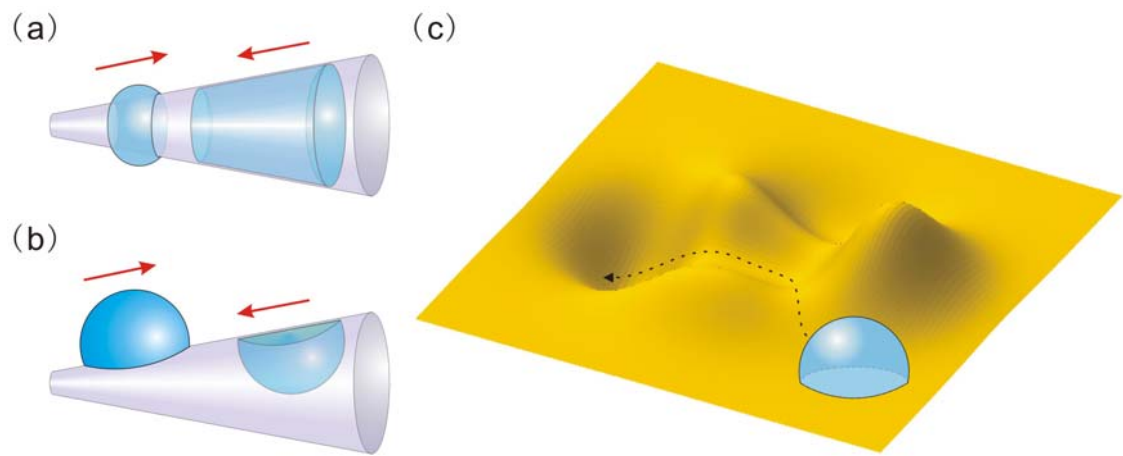

**Fig. 1** (a) Symmetrical water droplets deposited outside and inside of conical tube. (b) Asymmetrical water droplets deposited outside and inside of conical tube. (c) A droplet located on general curved surface with arbitrary curvature.

We notice that all the above researches are about annular droplets on conical surfaces (Fig. 1(a)). While smaller droplets deposited on conical surfaces with larger diameters tend to be clam-shell-shaped [10], as illustrated in Fig. 1(b). To the best of our knowledge, there is no report on self-motion of clam-shell-shaped droplets at a conical surface, or any other curved surface (Fig. 1(c)). Here we show that a nanoscale clam-shell-shaped water droplet at a graphene conical surface behaves similar as annular droplet.

## Methods

Molecular dynamics (MD) simulations were performed to study the water nanodroplet movement at the outer and inner surfaces of a graphene cone (Fig. 1(b)), based on the platform of LAMMPS [11]. In the simulation, the cone is treated as rigid and the SPC/E [12] model is employed to describe the water because of its better coincidence in surface tension 63.6 mN/m with experimental value 71.73 mN/m compared with other water models in room temperature [13]. Lennard-Jones potential is used as the water-carbon van der Waals interaction: $u(r) = 4\varepsilon[(\sigma/r)^{12} - (\sigma/r)^6]$. Here $u(r)$ is the potential between a pair of carbon-oxygen atoms, $r$ the distance between them, $\varepsilon$ the well depth of interaction; while the hydrogen-carbon interaction is neglected for the much weaker influence on the wetting property. The parameter $\sigma$ is related to the equilibrium distance of the carbon-oxygen pair.

To model different wettabilities, we fix $\sigma_{C-O} = 0.319$ nm and choose $\varepsilon_{C-O} = 5.8484$ meV, 1.9490 meV, and 1.50 meV, respectively, that correspond to water-graphene contact angles of 50.7° (hydrophilic), 138.1° (hydrophobic), and nearly 180° (superhydrophobic) [14]. Here we emphasize that the *graphene* is used here as a model, rather than a true material, in order to consider different wettabilities in an approch as simple as possible. In reality, the true contact angle of water at graphene is still an open problem.

First, we choose a cone with the half-apex angle of $\alpha = 19.5°$ and the height of 7 nm from the cone tip to the open end. Then, we cut off the tip of the cone at $Y = 1.5$ nm or $Y = 3.5$ nm for modeling the motion of a water droplet containing 339 atoms at the outer or inner conical surfaces, respectively. The diameter of a spherical droplet with 339 atoms is estimated about 2 nm.

In the simulation, the temperature of water is kept at 300 K with Nosé-Hoover thermostat [15, 16], and the whole system is located in a finite vacuum box. We first fix the initial mass center of droplet to apply thermal equilibrium for 100 ps. Then the droplet is released to move freely along the conical surface for next 300 ps. The trajectory results along the meridian line are illustrated in Fig. 2.

**Results and Discussions**

When the droplet was released at the outer surface, the simulation results show that the droplet immediately started a self-motion towards to the larger open end (*A*). After it had met the end *A*, it was bounced back, and then moved again towards the end *A*. The red solid curve in Fig. 2(a) demonstrates the moving distance along meridian line, *s*, versus simulation time, *t*, of the droplet at the outer surface with contact angle 50.8°, and the inserts are 5 selected frames from the movement movie (see Supplementary Information). Similar directional self-motion and bouncing were found for the hydrophobic (138.1°) and superhydrophobic (nearly 180°) surfaces, as shown by the black dashed- and blue dotted-lines in Fig. 2(a).

When the water droplet was released at the inner surface, we observed diametrically opposite results, as illustrated in the Fig. 2(b). The droplet started self-motion but towards to the smaller open end (*B*). Similar bouncing phenonomenon appeared when the droplet reaches the end *B*. The same tendency was found to be independent of wettability.

However, it can be seen from the motion curves in Fig. 2 that the smaller the contact angle of the surface has, the larger the self-motion speed of the droplet can reach. This rule is valid for movement at both outer and inner surfaces.

If we define the normal direction of the surface towards to droplet, then the two principal curvatures are $1/(s \tan\alpha)$ and 0 as the droplet sets at the outer surface of the cone, and $-1/(s \tan\alpha)$ and 0 as the droplet sets at the inner surfaces. Thus, from the above results one can easily conclude that the self-motion direction is always towards region with lower curvature.

The above findings, to the best of author's knowledge, are revealed and reported for the first time.

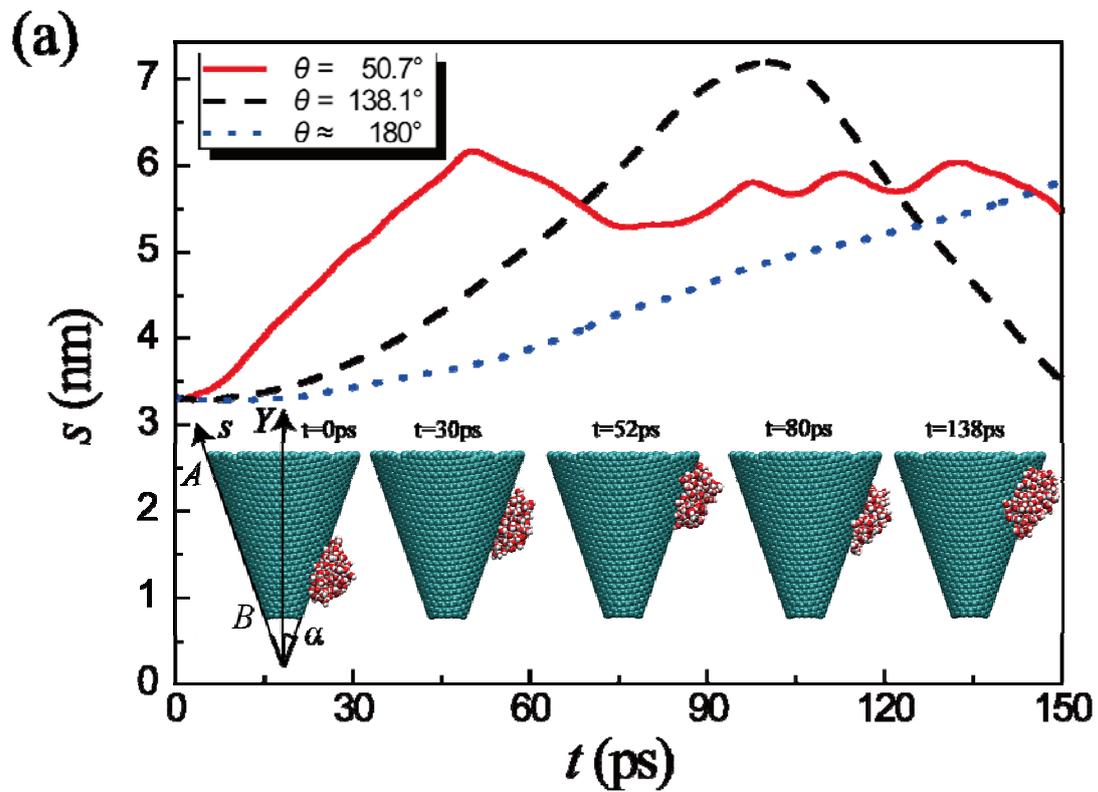

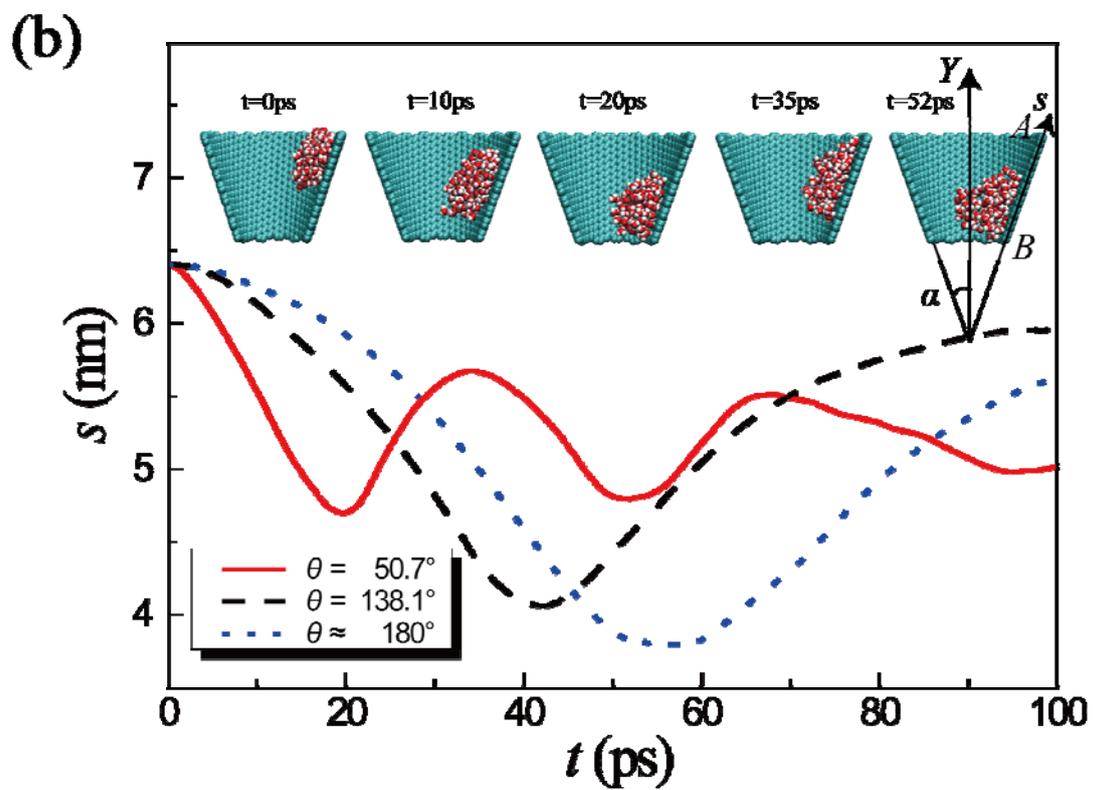

**Fig. 2** The wetting behaviors of a water droplet on the outer surface (a) and the inner surface (b) of the cone. The intrinsic Young's contact angles for red(solid), black(dash) and blue(dot) curves in both figures are 50.7°, 138.1° and nearly 180°, respectively. The configuration of the MD system is also illustrated in the insets of (a) and (b), where $\alpha$ is the half-apex angle, $s$-axis is parallel to the moving direction of the droplet and $Y$-axis is the central axis of carbon cone. Point $A$ (or $B$) denotes the upper (or lower) rim of the cone.

## Potential of single water molecule at a curved surface

To understand the above reported self-motion phenomenon, we study the van der Waals potential energy of a single water molecule at a curved surface. For simplicity, we first calculate the potential of a single water molecule at the conical outer and inner surfaces as function of the location. We move the water molecule along the meridian line of the conical surface, while keep the distance between the oxygen and the meridian a constant, $d = 0.5$ nm (see the inset in Fig. 3). The potential of water molecule and surface is obtained by adding all of the interaction between oxygen and carbon atoms. Here the interaction between oxygen and carbon atom is discribed by a typical Lennard-Jones parameters: $\sigma_{C-O} = 0.319$ nm, $\varepsilon_{C-O} = 5.8484$ meV, whose intrinsic contact angle is 50.7°.[14]

The results are plot in Fig. 3 as black upper triangles. We can see clearly that the potential of the single water molecule, predicted by this numerical method, increases along with the increase of the curvature. Thus the atom will spontaneously move from point $B$ to point $A$ at the outer surface, and from point $A$ to point $B$ at the inner surface, which is consistent completely with the moving tendencies of the global droplet in the previous section.

To reveal the influences of the curvature, we investigate the interactions between the water molecule and other three cones with different half-apex angles, i.e. $\alpha = 41.8°$, 30.0° and 9.6°, and the same contact angle 50.7° . The calculated potential-curvature relations are plotted by the other types of dots in Fig. 3, showing perfect coincidence with each other. This result verifies that the potential of the water molecule is determined by the curvature of the cone, while independent of the cone half-apex angle.

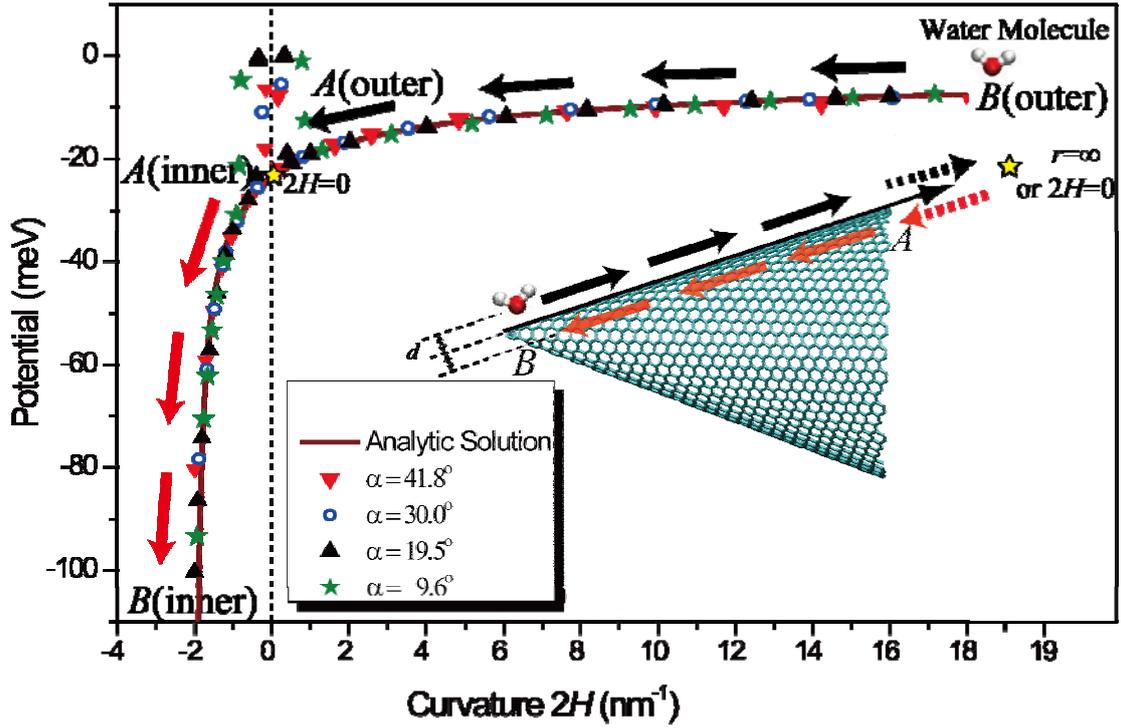

**Fig. 3** Relation between the potential of a single water molecule and the curvature (2*H*) of the nano-cone. The four discrete points sets present four different half-apex angles, and the solid curve is the result of analytical solution. The inset is the schemtic of the movement path for the water molecule, while the black and red arrows present that the moving path of molecule at the outer or inner surfaces of the cone, respectively.

Further and finally, we show that the above dependence of the potential on curvature is generally valid. After some derivations in the Appendix, the van der Waals potential of a single water molecule at a curved surface can be approximated in the following form:

$$U = U_0 \left(1 + 2Hd + Kd^2\right)^{-1/2} \tag{1}$$

where $U_0$ is the potential of the molecule with respect to a flat graphene, *H* and *K* are the mean and Gaussian curvatures of the surface, respectively. The driving force of the self-motion is thus in the inverse direction of gradient of this potential.

Geometrically, the factor $1+2Hd+Kd^2$ is the projection area of the parallel surface at the distance *d* where water molecule located to the unit area of the graphene surface, as illustrated in Fig. 4. The former outlines positions of the water molecules at contact with the graphene surface. The above explanation may help to obtain a geometrical understanding of the self-motion direction.

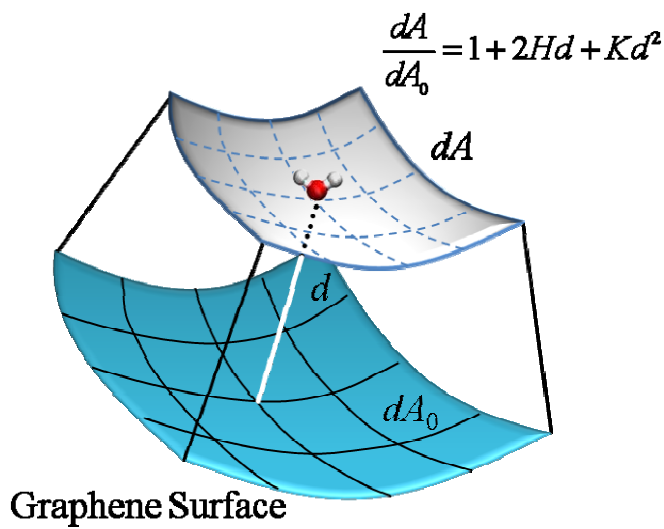

**Fig. 4** The parallel surface where water molecule located with graphene surface.

## Conclusions

In this paper, self-motion nanoscale water droplets and moving tendencies of water molecule on curved surfaces are studied through molecule dynamics simulations and analytical methods. We reveal that both the water droplet and single water molecule can move spontaneously from higher curvature regions to lower ones, independent of the wetting properties. The driving force comes from the curvature of the contacted curved surface. This result may be reagrded as the microscopic driving force mechanism for some dynamic phenomena occured on curved surfces, and provides a support to the opinion that the curvatures of curved spaces could induce driving forces [17].

## Acknowledgements

Financial support from the NSFC under grant No.10872114, No.10672089 and No. 10832005 is gratefully acknowledged.

**References**

1. Linke, H.; Aleman, B. J.; Melling, L. D.; Taormina, M. J.; Francis, M. J.; Dow-Hygelund, C. C.; Narayanan, V.; Taylor, R. P.; Stout, A., Self-propelled Leidenfrost droplets. *Phys. Rev. Lett.* **2006,** 96, 154502.
2. Sumino, Y.; Magome, N.; Hamada, T.; Yoshikawa, K., Self-running droplet: Emergence of regular motion from nonequilibrium noise. *Phys. Rev. Lett.* **2005,** 94, 068301.
3. Daniel, S.; Sircar, S.; Gliem, J.; Chaudhury, M. K., Ratcheting motion of liquid drops on gradient surfaces. *Langmuir* **2004,** 20, 4085-4092.
4. Kunihiro Ichimura, S.-K. O., Masaru Nakagawa, Light-driven motion of liquids on a photoresponsive surface. *Science* **2000,** 288, 1624-1626.
5. Prakash, M.; Quere, D.; Bush, J. W. M., Surface tension transport of prey by feeding shorebirds: The capillary ratchet. *Science* **2008,** 320, (5878), 931-934.
6. Lorenceau, E.; Quere, D., Drops on a conical wire. *J. Fluid Mech.* **2004,** 510, 29-45.
7. Liu, J.; Xia, R.; Li, B.; Feng, X., Directional motion of droplets in a conical tube or on a conical fibre. *Chin. Phys. Lett.* **2007,** 24, (11), 3210-3213.
8. Renvoise, P.; Bush, J. W. M.; Prakash, M.; Quere, D., Drop propulsion in tapered tubes. *Epl* **2009,** 86, (6), 5.
9. Zheng, Y.; Bai, H.; Huang, Z.; Tian, X.; Nie, F.-Q.; Zhao, Y.; Zhai, J.; Jiang, L., Directional water collection on wetted spider silk. *Nature* **2010,** 463, 640-643.
10. McHale, G.; Newton, M. I.; Carroll, B. J., The shape and stability of small liquid drops on fibers. *Oil & gas sci. tech. - Rev.IFP* **2001,** 56, (1), 47-54.
11. Plimpton, S., Fast parallel algorithms for short-range molecular-dynamics. *J. Comp. Phys.* **1995,** 117, (1), 1-19.
12. Berendsen, H. J. C.; Grigera, J. R.; Straatsma, T. P., The missing term in effective pair potentials. *J. Phys. Chem.* **1987,** 91, (24), 6269-6271.
13. Vega, C.; de Miguel, E., Surface tension of the most popular models of water by using the test-area simulation method. *J. Chem. Phys.* **2007,** 126, (15).
14. Werder, T.; Walther, J. H.; Jaffe, R. L.; Halicioglu, T.; Koumoutsakos, P., On the water-carbon interaction for use in molecular dynamics simulations of graphite and carbon nanotubes. *J. Phys. Chem. B* **2003,** 107, 1345-1352.
15. Nose, S., A molecular-dynamics method for simulations in the canonical ensemble. *Mol. Phys.* **1984,** 52, (2), 255-268.
16. Hoover, W. G., Canonical dynamics - Equilibrium phase-space distributions. *Phys. Rev. A* **1985,** 31, (3), 1695-1697.
17. Yin, Y. J.; Yin, J.; Ni, D., General mathematical frame for open or closed biomembranes (Part I): Equilibrium theory and geometrically constraint equation. *J. Math. Biol.* **2005,** 51, (4), 403-413.

# Appendix: Lennard-Jones potential between single molecule and general curved surface

A more general case, i.e. the interaction between the molecule and the general curved surface (Fig. A-1), is considered. Here point $O$ is the origin of the local coordinate system, located on the curved surface. $Ox$ and $Oy$ are tangential to the principle curvature lines, and $Oz$ is the outer normal of the curved surface at point $O$. The oxygen atom is fixed on the $z$ axis with distance $d$ from the point $O$.

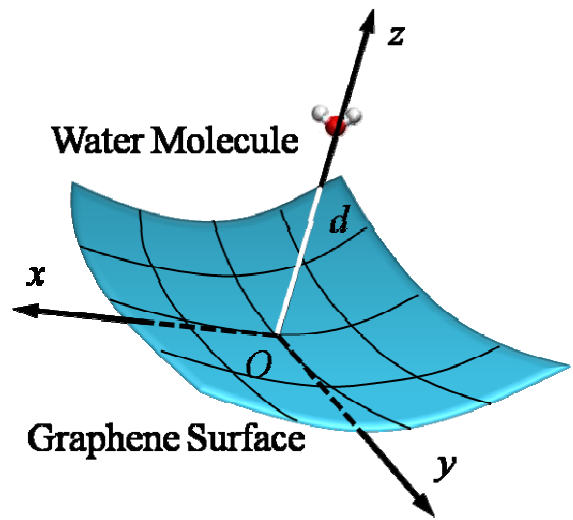

**Fig. A-1** The local cartansian coordinate system to depict the interaction between molecule and general curved surface, with $x$ axis and $y$ axis tangential respectively to the principle curvature lines at the point $O$.

The curved surface is described by $z = f(x, y)$ under the local coordinate system. Suppose the density of surface atoms, i.e. the number of surface atoms per unite area, is $\rho$. The effective interaction range of the Lennard-Jones potential is only about 1 nm, which means that the potential of the molecule is only affected by the atoms of surface in the neighborhood of molecule. Suppose that these surface atoms are distributed in the region with the characteristic radius $\delta$ around the center point $O$. Then the potential between molecule and surface atoms is:

$$U = \iint_{\substack{|x|<\delta \\ |y|<\delta}} 4\varepsilon \left[ \left(\frac{\sigma}{r}\right)^{12} - \left(\frac{\sigma}{r}\right)^{6} \right] \rho dS \tag{A-1}$$

For the expression, five procedures are adopted to make simplification as following:

(1) The area element $dS$. Considering that around the point $O$ there are approximations: $[\partial f(x,y)/\partial x]^2 \ll 1$ and $[\partial f(x,y)/\partial y]^2 \ll 1$, the area element $dS$ can be expressed as: $dS(x,y) = (1 + [\partial f(x,y)/\partial x]^2 + [\partial f(x,y)/\partial y]^2)^{1/2} dxdy \approx dxdy$.

(2) The surface function $f(x,y)$. Since the point $O$ is the origin point and $O$-$xy$ forms the tangential plane, we have these relationship: $f(0,0) = 0$, $\partial f(0,0)/\partial x = 0$ and $\partial f(0,0)/\partial x = 0$. And under the principle curvature system, we further get: $\partial^2 f(0,0)/\partial x \partial y = 0$. Based on these relationship, the second order Taylor expansion of surface function $f(x,y)$ can be simplified as: $f(x,y) \approx [\partial^2 f(0,0)/\partial x^2]x^2/2 + [\partial^2 f(0,0)/\partial y^2]y^2/2$.

(3) The distance $r$ between molecule and the curved surface. Since $f \ll d$ is valid around the $O$ point, we take one order approximation: $r^2 = x^2+y^2+[d-f(x,y)]^2 \approx x^2+y^2+d^2-2f(x,y)d$.

(4) The integral domain. Considering the far field only has weak influences, for simplification the domain of integration in Eq.(1) is extended into the whole surface, i.e. $|x| < \infty$ and $|y| < \infty$.

(5) The expression of curvatures at the point $O$. In the definition of coordinate system, we can get the mean curvature: $2H = -[\partial^2 f(0,0)/\partial x^2 + \partial^2 f(0,0)/\partial y^2]$ and Gauss curvature: $K = [\partial^2 f(0,0)/\partial x^2][\partial^2 f(0,0)/\partial y^2]$.

Integrating Eq. (A-1) with the five simplification leads to:

$$U = U_0 \left(1 + 2Hd + Kd^2\right)^{-1/2} \tag{1}$$

where $U_0 = 2\pi\rho\varepsilon\sigma^6[0.4(\sigma/d)^6-1]/d^4$.